# Fonts design – shapes processing of text information structures in process of non-invasive data acquisition

Ireneusz Kubiak
Military Communication Institute, Zegrze, Warszawska 22A St., Poland
e-mail: i.kubiak@wil.waw.pl

**Abstract.** *Computer fonts can be one of solutions supporting a protection of information against electromagnetic penetration. This solution is called „Soft TEMPEST". However, not every font has features which counteract the process of electromagnetic infiltration. The distinctive features of characters of font determine it. This article presents two sets of new computer fonts. These fonts are fully usable in everyday work. Simultaneously they make it impossible to obtain information using the non-invasive method. Names of these fonts are directly related to the shapes of the characters. Each character of these fonts is built only with vertical and horizontal lines. The differences between them consist in the different widths of the vertical lines. The Symmetrical Safe font is built from vertical lines with the same widths. The Asymmetrical Safe font is built from vertical lines with two different widths of lines. However, the appropriate proportions of the widths of the lines and clearances of each character of the safe font have to be met. The aim of this article is the presentation new solution in area of protection of information against electromagnetic penetration. It is a new approach which could replace old solutions in form of heavy shielding's, power and signal filters and also electromagnetic gaskets. Simultaneously, an application of new fonts is very easy. Only need to that the Asymmetrical Safe font or the Symmetrical Safe font is installed on computer station which processes text data.*

## 1. INTRODUCTION

An important element of daily processing of text information is an use of computers and computer fonts. There are many different computer fonts for a variety of applications. Most often, however, fonts are used to process text information, which is displayed on computer screens and printed by laser printers. Arial and Times New Roman fonts are the most popular. With their use, almost every text document is processed. The characters of these fonts have decorative elements such as: an ear, a bowl, an eya, a serif, a tail, a terminal, a bracket, a loop, etc. [1]. The characters are oval and also angles between the individual components are different than 90°. In addition, the widths of lines building characters are variable [2], especially for the Times New Roman font.

During the processing of text data, each character of font has its representation in the form of electrical signal. This signal is transmitted from a computer to a screen or to a laser printer. In this case, this signal become a source of electromagnetic emissions which have characteristics of this signal [3, 4, 5]. This phenomenon makes it possible to reconstruct every processed character and thus all text displayed on the screen or printed on the laser printer.

A Side Channel Attack (SCA) plays an important role in this process. SCA is created between the source of emission (electrical signal transmitted by e.g. video cables in VGA (Video Graphics Array, still very popular in non-public information systems, where first of all the text information is processed), DVI (Digital Visual Interface, HDMI – High Definition Multimedia Interface) or DisplayPort standards) and antenna which receives such emissions (Fig.1). These emissions are called revealing (sensitive) emissions [6, 7, 8].

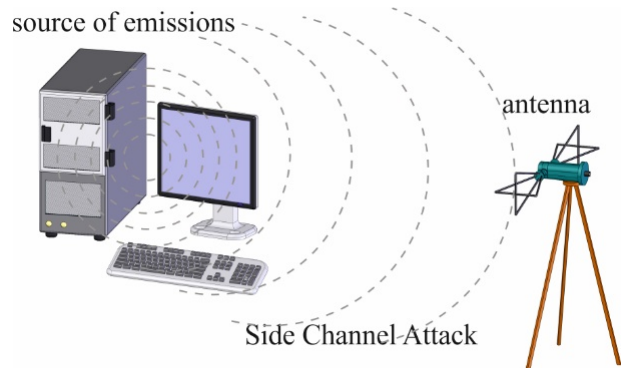

**Fig.1.** A Side Channel Attack

This type of SCA has the characteristics of a high-pass filter, which is an important property from the protection of information against electromagnetic infiltration process point of view. On the output of SCA only vertical and diagonal edges (rising edges and falling adges of pulses of electrical video signals) are visible on reconstructed images. There are no visible horizontal edges. In the case of the analog VGA standard, the reconstructed characters of the fonts are marked by their vertical and diagonal edges. For the digital DVI (HDMI) standard, due to the nature of the TMDS (Transition Minimized Differential Signaling) coding algorithm, there appear fills of characters. This phenomenon is related to the bit character (0 and 1) of the electrical signal and it may result in the loss of processed information (Fig.2) [9, 10].

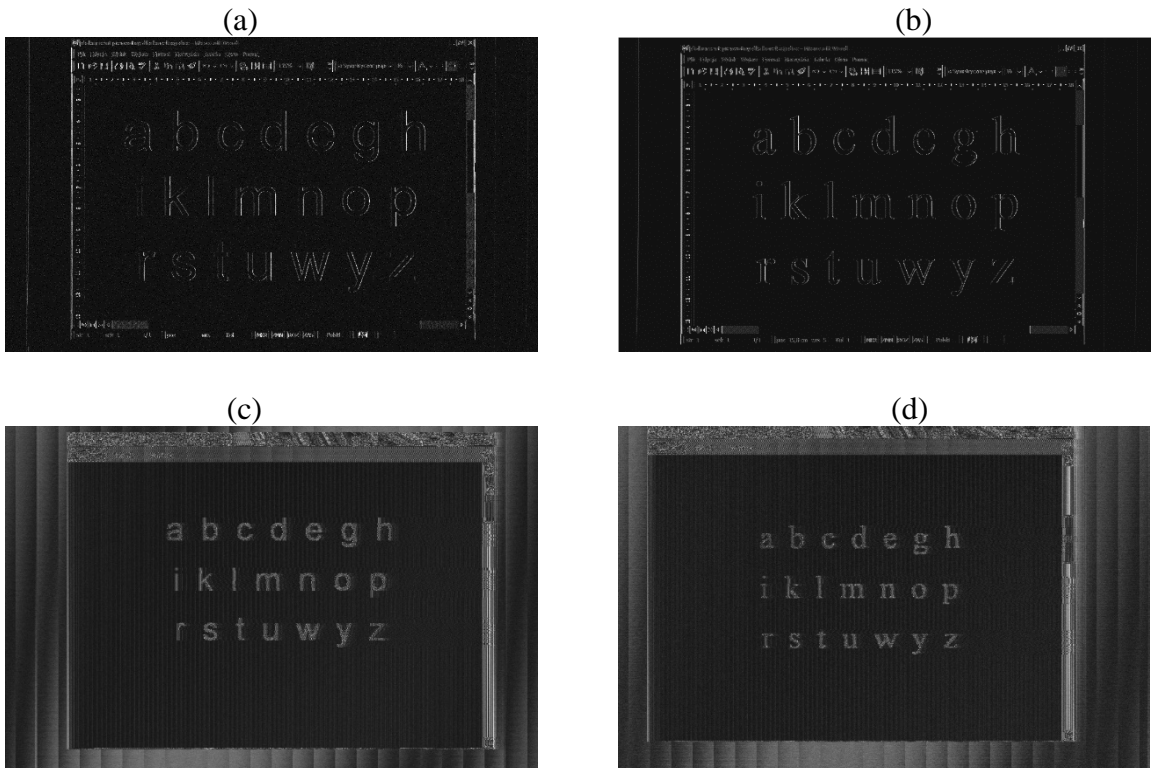

**Fig.2.** Examples of possibilities acquisition of information from sensitive emissions for text data and VGA and DVI (HDMI) standards: (a) Arial and VGA standard, (b) Times New Roman and VGA standard, (c) Arial and DVI (HDMI) standard, (d) Times New Roman and DVI (HDMI) standard

Another feature of the computer fonts is the possibility of transferring the paper form of a text document to an electronic version (editable e.g. in MS Word), using Optical Character Recognition (OCR) programs. Not every document should be moved to an editable form. Appropriate characters of fonts should counteract this process.

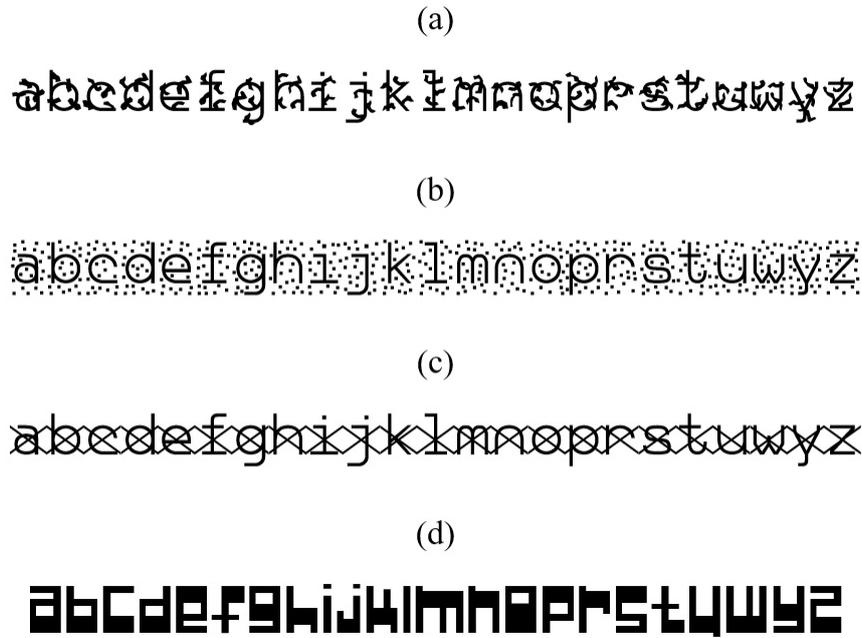

**Fig.3.** Sang Mun's fonts: (a) Flower, (b) Noise , (c) Cross and (d) Null Pointer font

Sang Mun tried to design such fonts. He created three sets fonts: Flower, Noise and Cross (Fig.3). Each font fulfils its role and counteracts optical characters recognition process. We also can find a font which has similar features as safe fonts. It is Null Pointer font (Fig.3d). Characters of this font have not diagonal lines but they have distinctive features. These features (different widths of vertical and horizontal lines) allow to recognize each character during the electromagnetic penetration process [11].

The safe fonts have both features. They are resistant to electromagnetic infiltration process [12, 13] as well as optical characters recognition process.

## 2. PROPOSED METHOD
### 2.1. Construction of characters of safe fonts
*2.1.1. Symmetrical Safe font*

The Symmetrical Safe font is one of two computer fonts which was designed according to safety criteria's [14, 15]. This font could be used in printing process and computer techniques. It is a typical font which is used in the processing of text data (technical documentation, scientific documentation, advertising banners), in particular of classified data. The font characters are devoid of decorative and diagonal elements. The lines building the characters intersect at a right angle. The general contour of characters of Symmetrical Safe font has a rectangle shape. Each character is built from lines about two widths. Wider lines are vertical lines of the character, thinner lines are horizontal lines of the character. Simultaneously the right proportions of the line width and the clearance of each character of the font are maintained. The corresponding characters have ascender and descender. There aren't unnecessary decorative elements. This makes font characters of font are similar each other with high values of correlation coefficient between characters. Examples of lowercase and digits are shown in Fig.4.

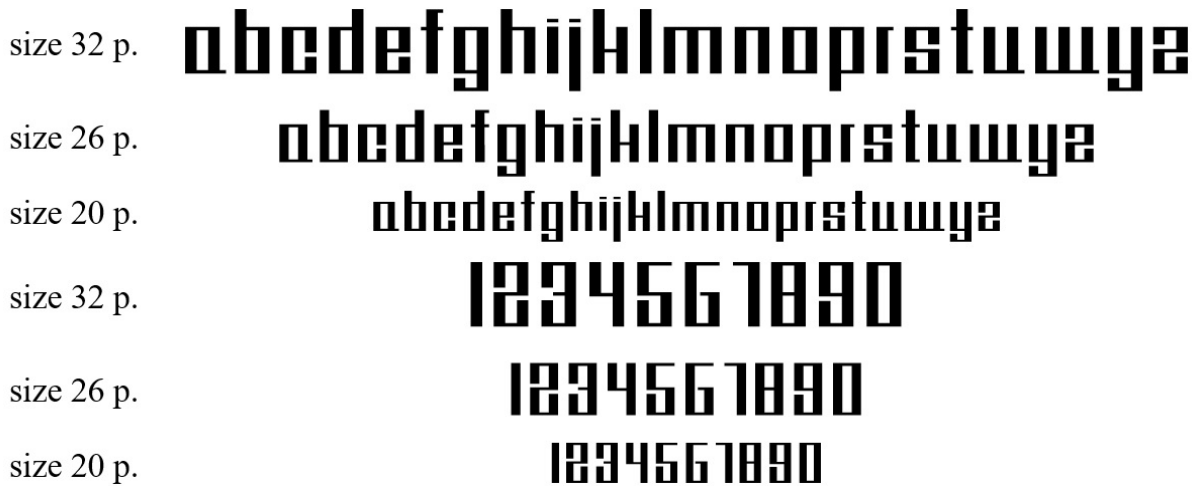

| | |
|---|---|
| size 32 p. | abcdefghijklmnoprstuwyz |
| size 26 p. | abcdefghijklmnoprstuwyz |
| size 20 p. | abcdefghijklmnoprstuwyz |
| size 32 p. | 1234567890 |
| size 26 p. | 1234567890 |
| size 20 p. | 1234567890 |

**Fig.4.** Characters of Symmetrical Safe font

The construction of characters, for ensure the counteracting properties of the electromagnetic eavesdropping process, has to meet appropriate assumptions. They concern the proportions of width of lines, clearance and the position of middle horizontal line. Such the line appears, for example, in the letters „e", „s" and „z" (Fig. 7). Details of construction of the characters of font are shown in Fig.5 and 6. They contain basic data regarding the width of characters, their height, as well as the ascender and descender. An important element of construction, affecting the readability of signs, is also appropriate space between the lines.

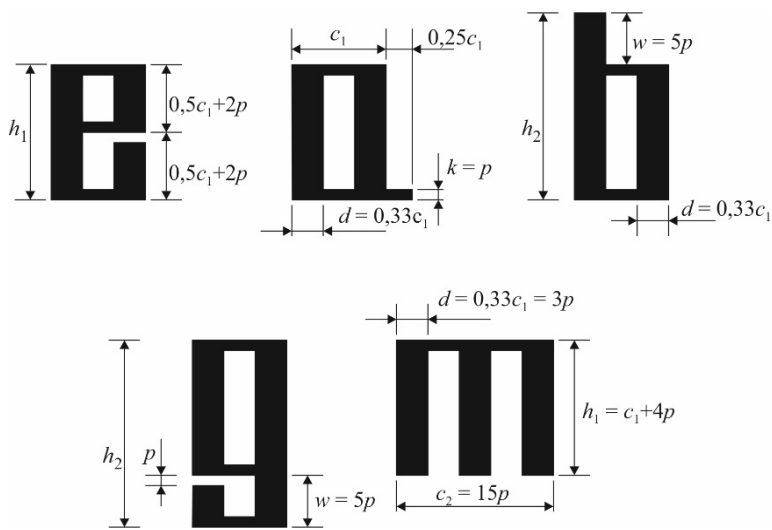

**Fig.5.** Construction of lowercase of Symmetrical Safe font

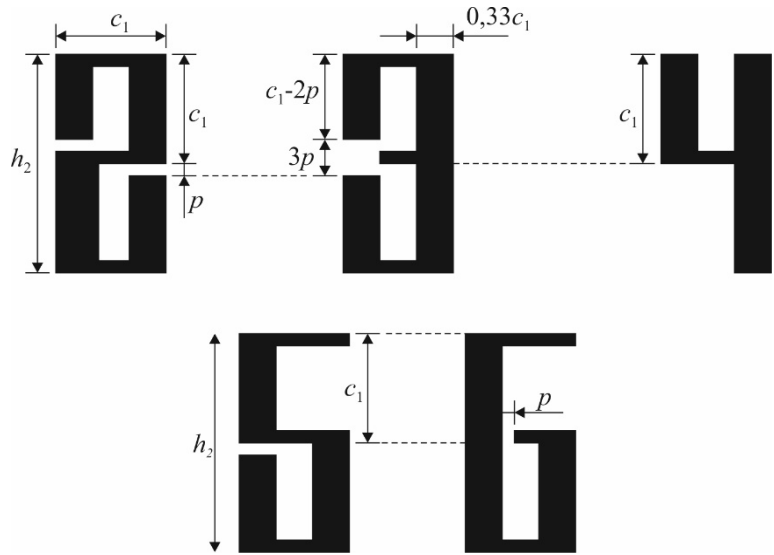

**Fig.6.** Construction of digits of Symmetrical Safe font

The characters of the font are constructed with the required proportions of the width of the lines to the clearances and the appropriate position of the middle horizontal lines for characters which have such lines. This applies to both lowercase and capital letters and also digits (Fig.7).

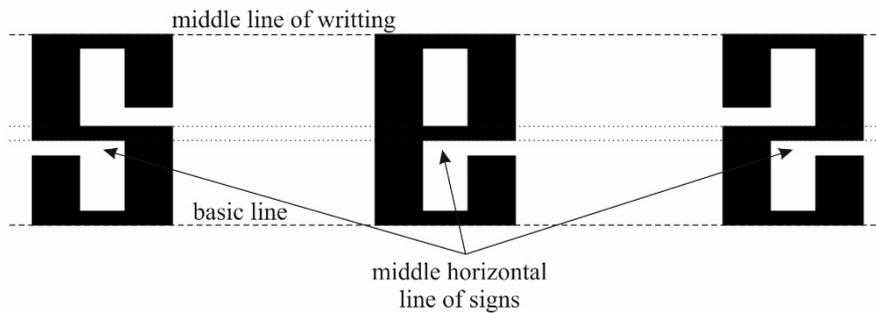

**Fig.7.** Location of middle horizontal line for lowercase of Symmetrical Safe font

Construction parameters of characters of Symmetrical Safe font:
a) height of lowercase:
$$h_1 = 13p, \qquad (1)$$
b) height character with ascender or descender (height of digits and capital letters):
$$h_2 = 18p, \qquad (2)$$
c) width of lowercase, capital letters and digits (with the exception of „m" character):
$$c_1 = 9p, \qquad (3)$$
d) width of „m" character (lowercase and capital letter):
$$c_2 = 15p, \qquad (4)$$
e) width of vertical line and clearance between vertical lines:
$$d = 3p, \qquad (5)$$
f) width of horizontal line:
$$k = p, \qquad (6)$$
g) ascender and descender:
$$w = 5p. \qquad (7)$$

*2.1.2. Asymmetrical Safe font*

The Asymmetrical Safe font [14, 15] is the second new computer. As the Symmetrical Safe font this font could be used in printing process and computer techniques. The characters of the font are devoid of decorative and diagonal elements. The lines building the characters intersect at a right angle. The general contour of characters of the Symmetrical Safe font has a rectangle shape. Each character is also built from lines about two widths. However, the location of the lines in the characters is different than for the Symmetrical Safe font. Wider lines are vertical lines but only as a left part of the character. Thinner lines appear as horizontal lines of the character and a right element of the character (Fig.8). Simultaneously the right proportions of width of the lines and the clearance of each character of font are maintained. The corresponding characters have ascender and descender.

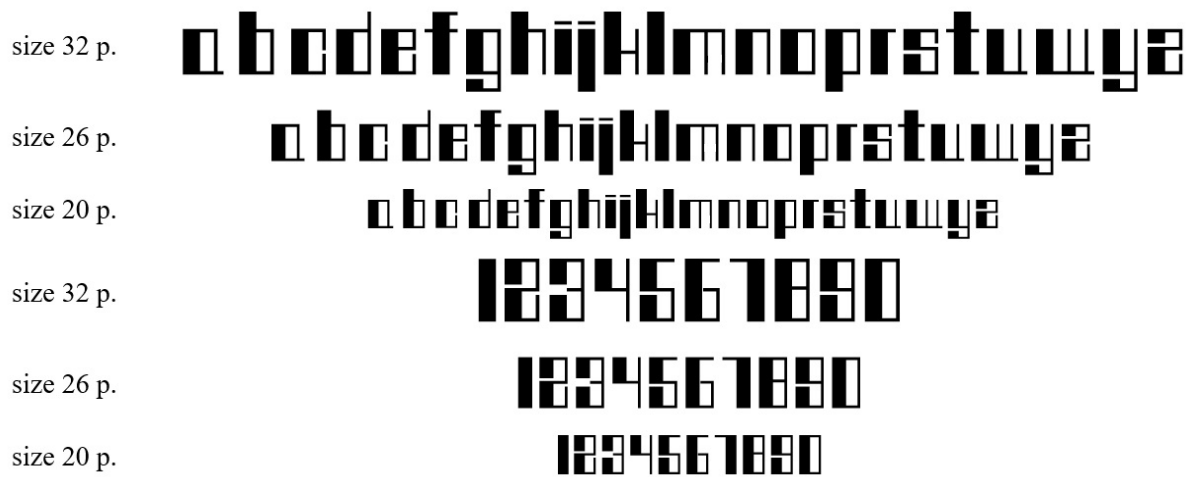

**Fig.8.** Characters of Asymmetrical Safe font

There haven't necessary decorative elements. This makes that characters of font are similar each other with high values of correlation coefficient between characters. Examples of lowercase and digits are shown in Fig.8.

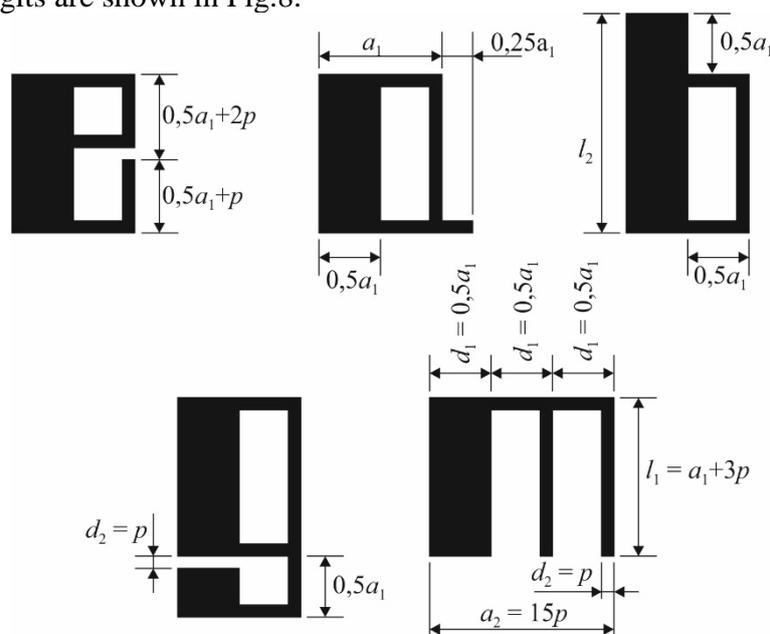

**Fig.9.** Construction of lowercase of Asymmetrical Safe font

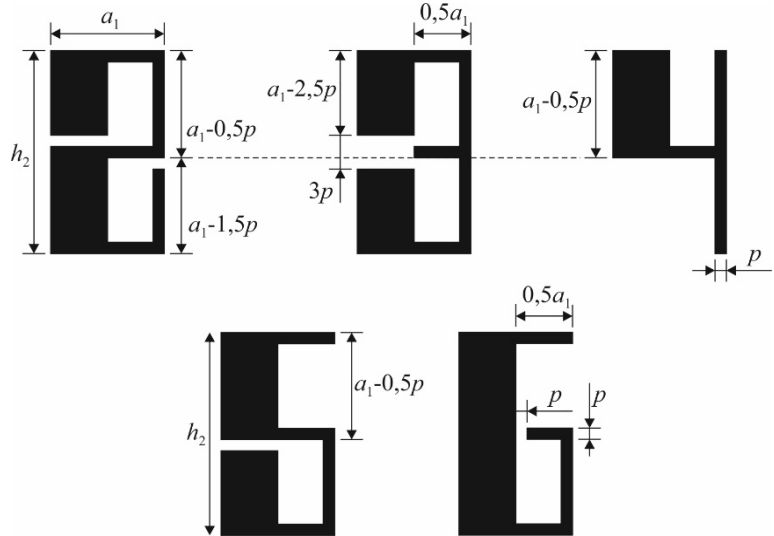

**Fig.10.** Construction of digits of Asymmetrical Safe font

Similarly to the Symmetrical Safe font, the construction of characters of the Asymmetrical Safe font has to meet the appropriate assumptions (Fig.9 and 10) to ensure the protection against the electromagnetic eavesdropping process. They concern the proportions of width of lines, clearance and the horizontal position of the middle line. Such the line appears, for example, in the letters „e", „s" and „z" (Fig.11).

The characters of font are constructed with the required proportions of the width of the lines to the clearances and the appropriate position of the middle horizontal line for characters which have such line. This applies to both lowercase and capital letters and also digits (Fig.11).

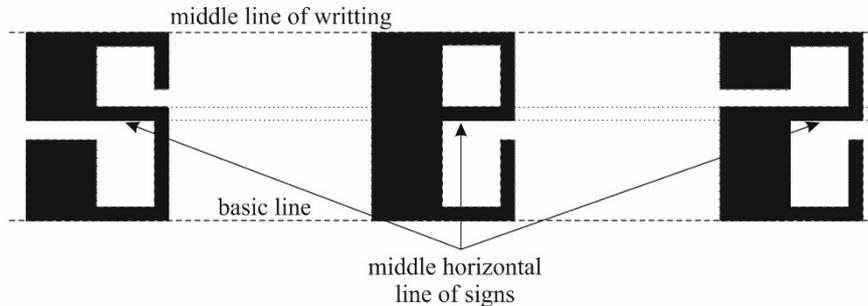

**Fig.11.** Location of middle horizontal line for lowercase of Asymmetrical Safe font

Construction parameters of characters of Asymmetrical Safe font:
a) height of lowercase:
$$h_1 = 13p, \qquad (8)$$
b) height character with ascender or descender (height of digits and capital letters):
$$h_2 = 18p, \qquad (9)$$
c) width of lowercase, capital letters and digits (with the exception of „m" character):
$$a_1 = 10p, \qquad (10)$$
d) width of „m" character (lowercase and capital letter):
$$a_2 = 15p, \qquad (11)$$
e) width of vertical lines and clearance between vertical lines:
$$d_1 = 5p, \quad d_2 = p \qquad (12)$$
f) width of horizontal lines:
$$k = p, \qquad (13)$$
g) ascender and descender:

$$w = \frac{1}{2}a_1 = 5p. \tag{14}$$

## 3. EXPERIMENTAL RESULTS AND DISCUSIONS
## 3.1. Electromagnetic protection as an application of safe fonts

The construction of characters of Symmetrical Safe and Asymmetrical Safe fonts has an effect on high level of similarity between the characters. Because SCA, due to its properties, additionally eliminates some construction elements of characters, an attempt of identification of characters becomes impossible [16, 17]. In reconstructed images from sensitive emissions, almost all characters (depends on the degree of similarity and correlation coefficient value) are identified as one and the same [18, 19, 20].

(a)

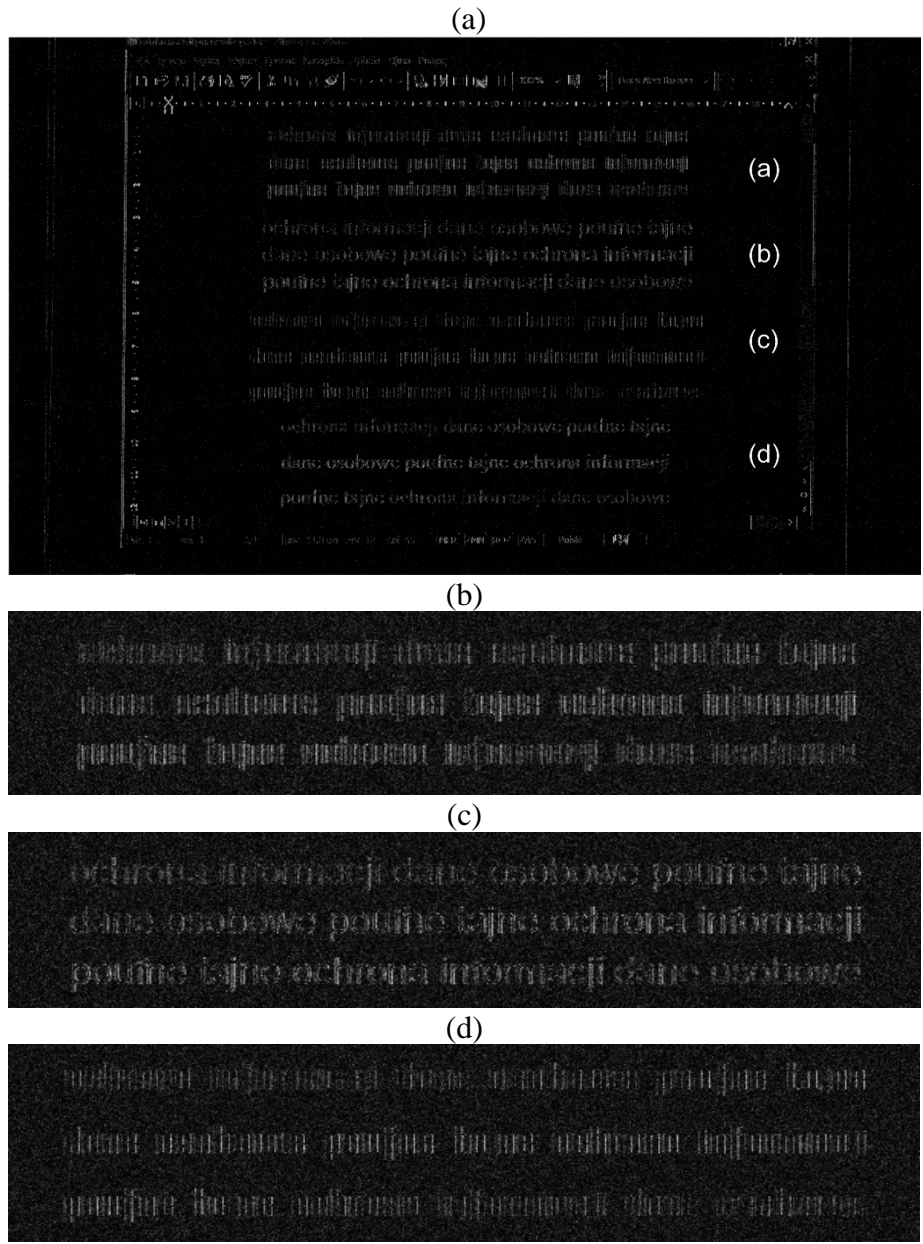

(b)

(c)

(d)

(e)

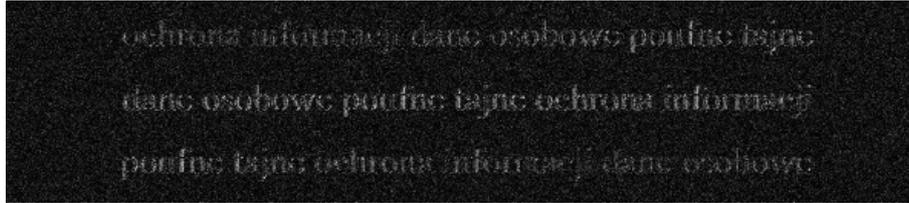

**Fig.12.** (a) Example of reconstructed image with text data for four different fonts (from the top: Symmetrical Safe, Arial, Asymmetrical Safe, Times New Roman) and magnifications of parts of the image: (b) Symmetrical Safe, (c) Arial, (d) Asymmetrical Safe, (e) Times New Roman (frequency of appearing of reveal emission: $f_0$ = 740 MHz, *BW* = 50 MHz, size of characters: 14 p., VGA standard was a source of reveal emission)

(a)

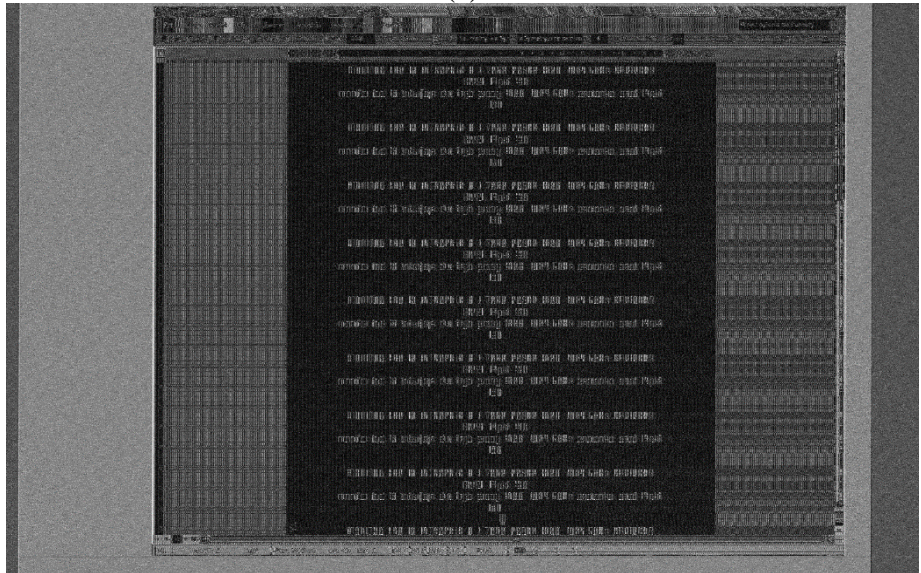

(b)

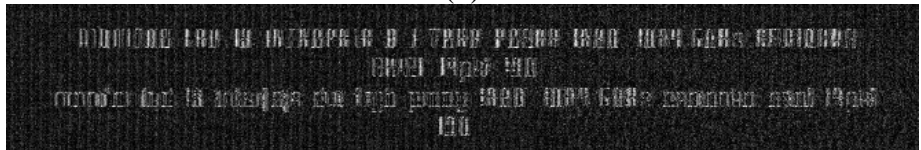

(c)

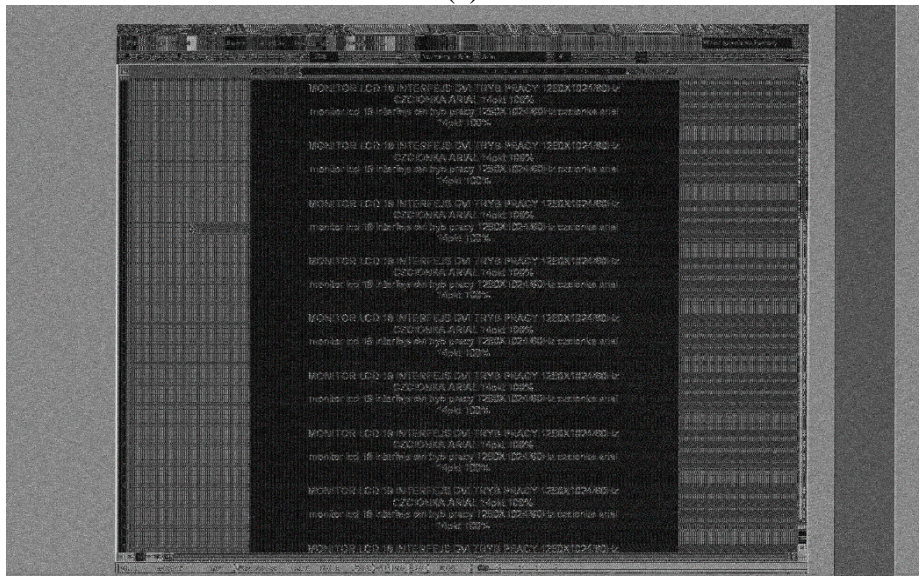

9(e)

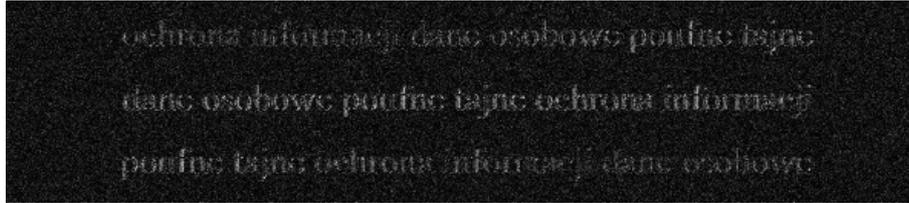

**Fig.12.** (a) Example of reconstructed image with text data for four different fonts (from the top: Symmetrical Safe, Arial, Asymmetrical Safe, Times New Roman) and magnifications of parts of the image: (b) Symmetrical Safe, (c) Arial, (d) Asymmetrical Safe, (e) Times New Roman (frequency of appearing of reveal emission: $f_0$ = 740 MHz, *BW* = 50 MHz, size of characters: 14 p., VGA standard was a source of reveal emission)

(a)

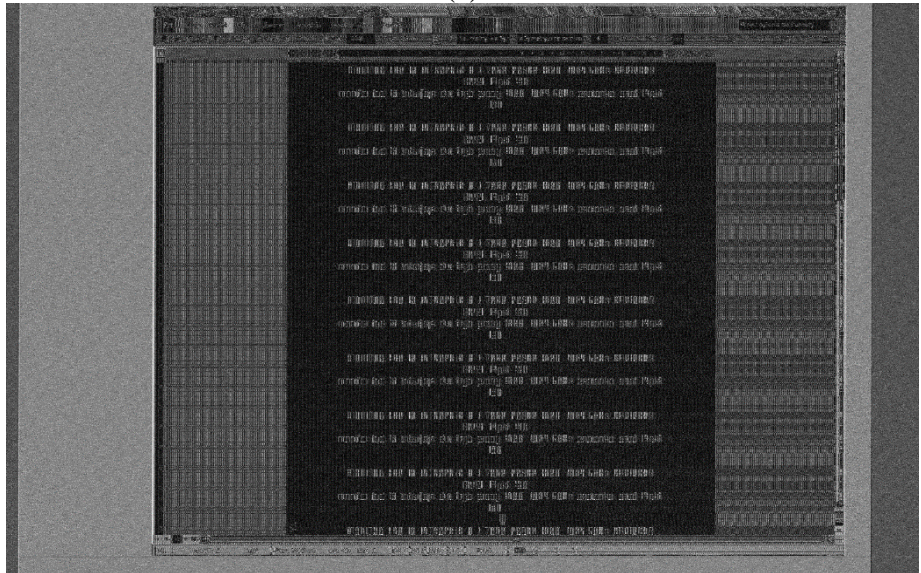

(b)

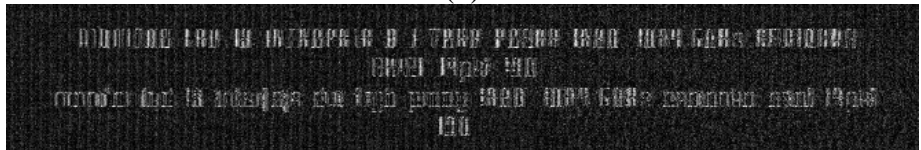

(c)

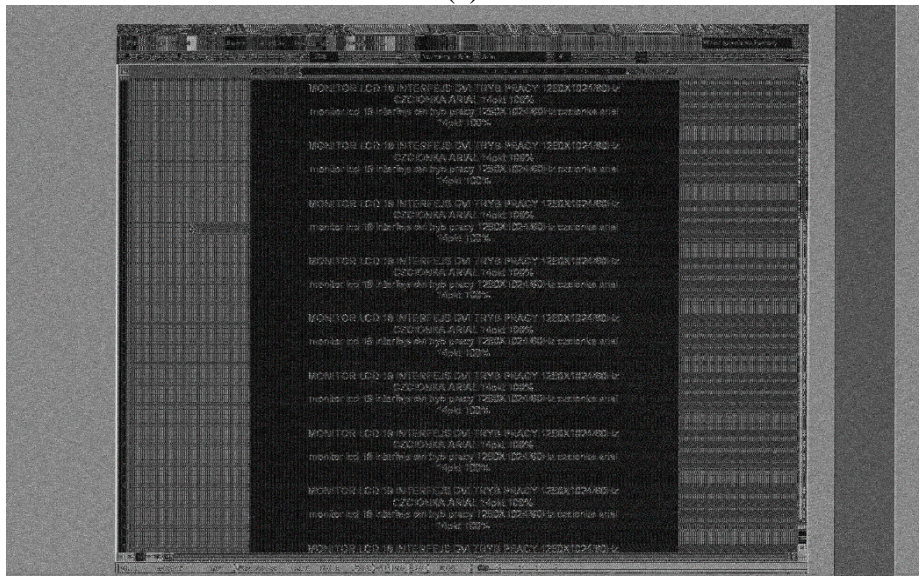

(d)

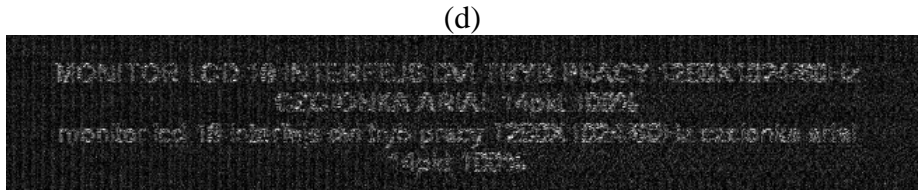

**Fig.13.** Examples of reconstructed images with text data: (a) Symmetrical Safe, (c) Arial, and magnifications of parts of these images ((b) and (d)) (frequency of appearing of reveal emission: $f_0$ = 1730 MHz, $BW$ = 100 MHz, size of characters: 14 p., DVI (HDMI) standard was a source of reveal emission)

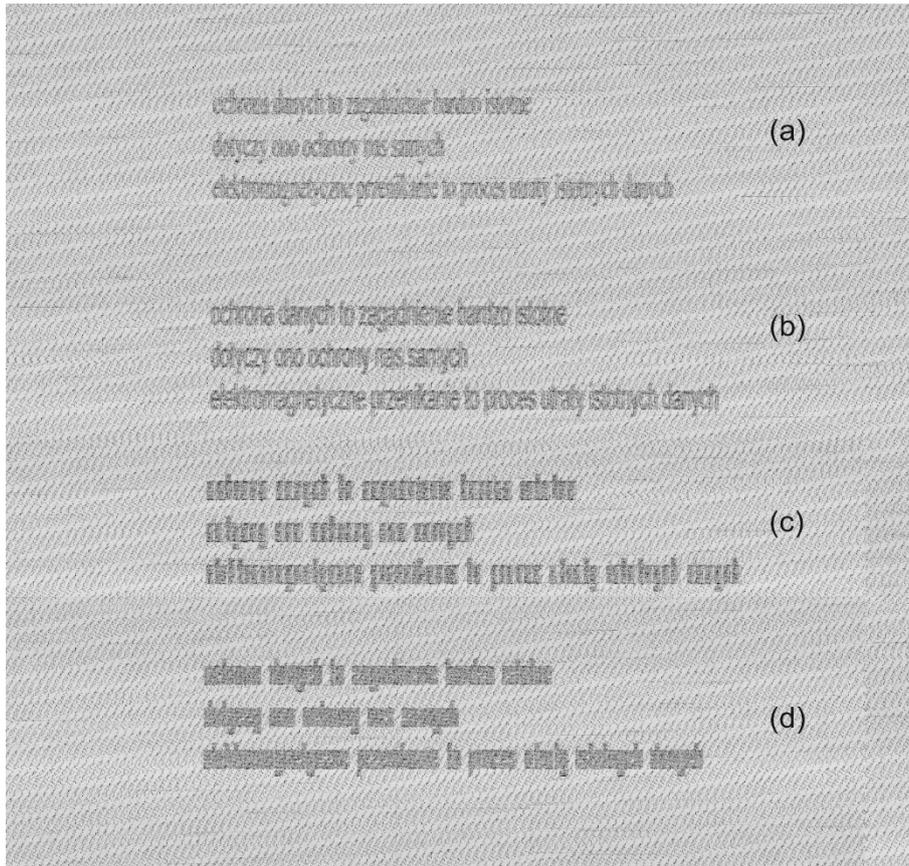

**Fig.14.** Example of reconstructed image with text data for four different fonts: (a) Times New Roman, (b) Arial, (c) Asymmetrical Safe, (d) Symmetrical Safe (frequency of appearing of reveal emission: $f_0$ = 750 MHz, $BW$ = 30 MHz, printing resolution (size of characters: 14 p.): 600 dpi × 600 dpi, laser printer was a source of reveal emission)

In this situation the reading of text data makes impossible (Fig.12, 13 and 14). This makes that safe fonts are useful not only in the typical processing of text data. These fonts effectively counteract the process of electromagnetic infiltration.

### 3.2. Character Error Rate

The effectiveness of safe fonts in the process of protection of information is confirmed by very large values of Character Error Rate (CER, Tab.1), which was counted according to below equation:

$$CER = \frac{m+k}{q} = \frac{m+(u-n)}{q}, \qquad (15)$$

where:

$u$ – the number of characters looked for in analysed image, $m$ – the number of characters incorrectly recognized, $n$ – the number of characters correctly recognized, $k$ – the number of unrecognized but looked for characters in analysed image ($k = u – n$), $q$ – the number of all characters existing in analysed image. This means that during reconstruction of characters of letters and digits there are made a lot of mistakes [21, 22, 23]. This applies to both lowercase and capital letters and also digits (Fig.12, 13 and 14).

**Table 1.** Values of Character Error Rate for traditional and safe fonts [24]

| Character | Arial font | Times New Roman font | Symmetrical Safe font | Asymmetrical Safe font |
|---|---|---|---|---|
| *VGA standard* | | | | |
| a | 1 | 2 | 1 | 4 |
| c | 3 | 5 | 7 | 350 |
| h | 51 | 0 | 310 | 6 |
| n | 31 | 39 | 283 | 183 |
| s | 1 | 3 | 40 | 204 |
| *DVI (HDMI) standard* | | | | |
| a | 1 | 4 | 91 | 21 |
| c | 3 | 11 | 97 | 509 |
| h | 3 | 4 | 50 | 60 |
| n | 7 | 6 | 431 | 80 |
| s | 1 | 1 | 156 | 70 |
| *Laser printer (double diode system, resolution: 600 dpi × 600 dpi)* | | | | |
| a | 3 | 5 | 3 | 122 |
| c | 0 | 5 | 19 | 98 |
| h | 5 | 0 | 20 | 95 |
| n | 3 | 10 | 73 | 104 |
| s | 2 | 3 | 3 | 75 |

## 4. CONCLUSIONS

The safe fonts are fully functional computer fonts and they can be used to process text information. A feature which distinguishes these fonts from traditional Arial or Times New Roman fonts is the high level of similarity between the characters of font. This property was achieved by eliminating decorative elements and diagonal lines, building characters of fonts. This means that characters of safe fonts have unique shapes which aren't similar to other existing characters of computer fonts.

The safe fonts via their unique characters are characterized by significant features related to protection against electromagnetic penetration process. In reconstructed images from sensitive emissions, there is difficult to indicate a specific character, in contrast to traditional fonts. This applies to sources of unwanted emissions in the form of analogue and digital graphic standards [25]. Additionally, the safe fonts are resistant to optical character recognition. Software of this type can not correctly recognize of letters and digits.

Symmetrical Safe and Asymmetrical Safe fonts are new computer fonts. Due to their universality of the use and acceptance of potential users, the works on improvement of the shapes of characters are still being continued. Despite of the high level of similarity between the characters of font, each safe font could be used in the secure processing of information. These fonts obtained protection of the Polish Office Pattern in the form of Industrial Design (No. 24487) [14] and Pattern (No. P.408372) [15].

**LITERATURE**

[1] A. Shainir, A. Rappoport, Extraction of Typographic Elements from Outline Representations of Fonts, *Computer Graphics Forum*, Volume 15, Issue 3, 1996, DOI: https://doi.org/10.1111/1467-8659.1530259

[2] Z. Nanxuan, C. Ying, W.H. Lau, Modeling Fonts in Context: Font Prediction on Web Designs, *Computer Graphics Forum*, Volume 37, Issue 7, 2018, https://doi.org/10.1111/cgf.13576

[3] S. Jun, A. Yongacoglu, D. Sun, W. Dong, Computer LCD recognition based on the compromising emanations in cyclic frequency domain, *IEEE International Symposium on Electromagnetic Compatibility*, 25-29 July 2016, Ottawa, Canada, pp.164-169, DOI: 10.1109/ISEMC.2016.7571637

[4] H.K. Lee, J.H. Kim, S.C. Kim, Emission Security Limits for Compromising Emanations Using Electromagnetic Emanation Security Channel Analysis, *IEICE Transactions on Communications*, 1 October 2013, pp. 2639-2649, https://doi.org/10.1587/transcom.E96.B.2639

[5] S. Tae-Lim, J. Yi-Ru, Y. Jong, Modeling of Leaked Digital Video Signal and Information Recovery Rate as a Function of SNR, *IEEE Transactions on Electromagnetic Compatibility*, Vol. 57, 2015, pp. 164-172

[6] I. Kubiak, Video signal level (colour intensity) and effectiveness of electromagnetic infiltration, *Bulletin of the Polish Academy of Sciences - Technical Sciences*, vol. 64, 2016, pp. 207-2018, doi: 10.1515/bpasts-2016-0023

[7] I. Kubiak, Laser printer as a source of sensitive emissions, *Turkish Journal of Electrical Engineering & Computer Sciences*, Vol. 26 No. 3, 2018, pp. 1354-1366, DOI: 10.3906/elk-1704-263

[8] Z. Mahshid, H.T. Saeedeh, G. Ayaz, Security limits for Electromagnetic Radiation from CRT Display, *Second International Conference on Computer and Electrical Engineering*, Dubai, United Arab Emirates, 28-30 January 2009, pp. 452-456

[9] Z. Nan, L. Yinghua, C. Qiang, W. Yiying, Investigation of Unintentional Video Emanations from a VGA Connector in the Desktop Computers, *IEEE Transactions on Electromagnetic Compatibility*, Vol.59, No 6, December 2017, pp. 1826-1834

[10] L. Dejiang, S. Jian, Y. Hongsheng, N. Qiang, G. Qingxi, Recognition and localization of actinidia arguta based on image recognition, *EURASIP Journal on Image and Video Processing,* 2019:21, 25 January 2019

[11] I. Kubiak, LED printers and safe fonts as an effective protection against the formation of unwanted emission, *Turkish Journal of Electrical Engineering & Computer Sciences*, Vol. 25, No. 5, 2017, pp. 4268-4279, DOI: 10.3906/elk-1701-128

[12] I. Kubiak, TEMPEST font counteracting a non-invasive acquisition of text data, *Turkish Journal of Electrical Engineering & Computer Sciences*, Vol. 26, No. 1, 2018, pp. 582-592, DOI: 10.3906/elk-1704-9

[13] I. Kubiak, Industrail Design No. 24487, Polish Pattern Office, 10 September 2018

[14] I. Kubiak, Pattern No. P.408372 Method for protecting transmission of information, Polish Pattern Office, 3 December 2018

[15] Y. Wenrui, Analysis of sports image detection technology based on machine learning, EURASIP Journal on Image and Video Processing, 2019:17, 21 January 2019

[16] X. Mingyuan, W. Yong, Research on image classification model based on deep convolution neural network, EURASIP Journal on Image and Video Processing 2019 2019:40, 11 February 2019

[17] J. Manson, S. Schaefer, Wavelet Rasterization, Computer Graphics Forum, Volume 30, Issue 2, 2011, https://doi.org/10.1111/j.1467-8659.2011.01887.x

[18] L. Jinming, Z. Jiemin, L. Taikang, L. Yongmei, The reconstitution of LCD compromising emanations based on wavelet denoising, 12th International Conference on Computer Science and Education (ICCSE), Houston, USA, pp. 294-297, 22-25 August 2017, 10.1109/ICCSE.2017.8085505

[19] C. Nidhi, G. Amnesh, A Technique for Image Encryption with Combination of Pixel Rearrangement Scheme Based On Sorting Group-Wise of RGB Values and Explosive Inter-Pixel Displacement, International Journal of Image, Graphics and Signal Processing, Vol.2, 2012, pp.16-22, DOI: 10.5815/ijigsp.2012.02.03

[20] W. Litao, Y. Bin, Analysis and Measurement on the Electromagnetic Compromising Emanations of Computer Keyboards, Seventh International Conference on Computational Intelligence and Security, 3-4 December 2011, Sanya, pp. 640-643, DOI: 10.1109/CIS.2011.146

[21] W. Litao, Y. Bin, Research on the compromising electromagnetic emanations from digital signals, International Conference on Automatic Control and Artificial Intelligence (ACAI 2012), January 2012, pp. 1761-1764, DOI: 10.1049/cp.2012.1329

[22] I. Kubiak, The Influence of the Structure of Useful Signal on the Efficacy of Sensitive Emission of Laser Printers, Measurement, Vol. 119, April 2018, pp.63-76, https://doi.org/10.1016/j.measurement.2018.01.055

[23] I. Kubiak, Computer font resistant to electromagnetic infiltration, Publisher House of Military University of Technology, 2014, ISBN 978-83-7938-018-3

[24] A.B. Tokarev, V.M. Pitolin, S.Y. Beletskaya, A.V Bulgakov, Detection of informative components of compromising electromagnetic emanations of computer hardware, International Journal of Computer Technology and Applications, Vol.9, January 2016, pp. 09-19